\documentclass{WileyMSP-template}

\usepackage{amsfonts}
\usepackage[print-unity-mantissa=false, separate-uncertainty=true]{siunitx}
\usepackage{physics}
\usepackage{booktabs}
\usepackage{subcaption}
\usepackage{textalpha}
\usepackage{isomath}
\usepackage{hyperref}

\AtBeginDocument{\RenewCommandCopy\qty\SI}

\usepackage{scalerel}
\newcommand{\lt}{\ell^\ast}
\newcommand{\ls}{\ell}
\newcommand{\lavg}{\expval{\ell}}

\newcommand{\lz}{\ell_z}
\newcommand{\musp}{\mu_\text{s}^\prime}
\newcommand{\muspi}{\mu_{\text{s}, i}^\prime}
\newcommand{\mus}{\mu_\text{s}}
\newcommand{\mua}{\mu_\text{a}}
\newcommand{\muext}{\mu_\text{ext}}
\newcommand{\mux}{\mu_{\text{s}, x}}
\newcommand{\muy}{\mu_{\text{s}, y}}
\newcommand{\muz}{\mu_{\text{s}, z}}
\newcommand{\mui}{\mu_{\text{s}, i}}
\newcommand{\muj}{\mu_{\text{s}, j}}
\newcommand{\muk}{\mu_{\text{s}, k}}
\newcommand{\ze}{z_\text{e}}
\newcommand{\Dtens}{\mathsfbfit{D}}
\newcommand{\mutens}{\mathsfbfit{\mus}}

\newcommand{\gtens}{\mathsfbfit{g}}
\newcommand{\Bze}{\mathcal{B}}
\newcommand{\Cze}{\mathcal{C}}
\newcommand{\Xze}{\mathcal{X}}
\newcommand{\Yze}{\mathcal{Y}}
\DeclareMathOperator{\sgn}{sgn}
\DeclareMathOperator{\acoth}{arccoth}

\newcommand{\piup}{\text{\textpi}}
\newcommand{\eu}{\ensuremath{\mathrm{e}}}

\usepackage{pgfplots}
\pgfplotsset{compat=newest}
\usepgfplotslibrary{external, fillbetween, colorbrewer, groupplots, units, polar}
\usetikzlibrary{external, shadows, arrows.meta, decorations.markings, fadings, patterns}
\pgfplotsset{compat=newest,
	/pgf/number format/1000 sep={\,},
	tick label style={font=\footnotesize},
	label style={font=\small},
	legend style={font=\footnotesize, draw=none, fill=none},
	height=6cm,
	width=\columnwidth,
	major grid style={lightgray, ultra thin},
	legend cell align={left},
	legend image code/.code={\draw[mark repeat=2,mark phase=2] plot coordinates {(0cm,0cm) (0.16cm,0cm) (0.32cm,0cm)};},
	filter discard warning=false,
	xlight/.style={Paired-E, semithick},
	xdark/.style={Paired-F, semithick},
	ylight/.style={Paired-C, semithick},
	ydark/.style={Paired-D, semithick},
}

\begin{document} 
	
\pagestyle{fancy}

\title{Generalized diffusion theory for radiative transfer in fully anisotropic scattering media}

\maketitle

\author{Ernesto Pini\textsuperscript{1,2,*,\S}}
\author{Michele Giusfredi\textsuperscript{3,4,5,*,\#}}
\author{Lorenzo Pattelli\textsuperscript{1,2,6,\dag}}

\begin{affiliations}
	\textsuperscript{1} Istituto Nazionale di Ricerca Metrologica (INRiM), 10135, Turin, Italy\\
	\textsuperscript{2} European Laboratory for Non-linear Spectroscopy (LENS), 50019, Sesto Fiorentino, Italy\\
	\textsuperscript{3} Department of Physics and Astronomy, Universit\`a di Firenze, 50019, Sesto Fiorentino, Italy\\
	\textsuperscript{4} Istituto dei Sistemi Complessi, Consiglio Nazionale delle Ricerche, 50019, Sesto Fiorentino, Italy\\
	\textsuperscript{5} Istituto Nazionale di Fisica Nucleare, Sezione di Firenze, 50019, Sesto Fiorentino, Italy\\
	\textsuperscript{6} National Research Council -- National Institute of Optics (CNR-INO), 50019, Sesto Fiorentino, Italy\\[6pt]
	
	\textsuperscript{*} These authors contributed equally to this work.\\
	\textsuperscript{\S} Ernesto Pini, Email: \url{pinie@lens.unifi.it}\\
	\textsuperscript{\#} Michele Giusfredi, Email: \url{michele.giusfredi@unifi.it}\\
	\textsuperscript{\dag} Corresponding author: Lorenzo Pattelli, Email: \url{l.pattelli@inrim.it}
\end{affiliations}


\begin{abstract}
A generalized anisotropic-diffusion framework is developed for transport problem in media described by a tensorial scattering coefficient and a scalar Henyey--Greenstein asymmetry factor.
In this regime the classical similarity relation between scattering and transport parameters fails, and each principal diffusion coefficient depends on all components of the microscopic scattering rate.
Explicit expressions are derived for the direction-averaged mean free path, the diagonal elements of the diffusion tensor, and boundary condition lengths via rapidly convergent spherical-harmonics expansions, along with open-source implementations.
The resulting predictions are validated against anisotropic Monte Carlo simulations, showing excellent agreement across broad ranges of structural anisotropy and phase-function asymmetry factors.
The theory provides a compact, general route connecting microscopic anisotropic scattering to macroscopic diffusion coefficients and boundary conditions in bounded geometries.
\end{abstract}
	
\section{Introduction}

\begin{figure}
	\centering
	\includegraphics[width=0.6\textwidth]{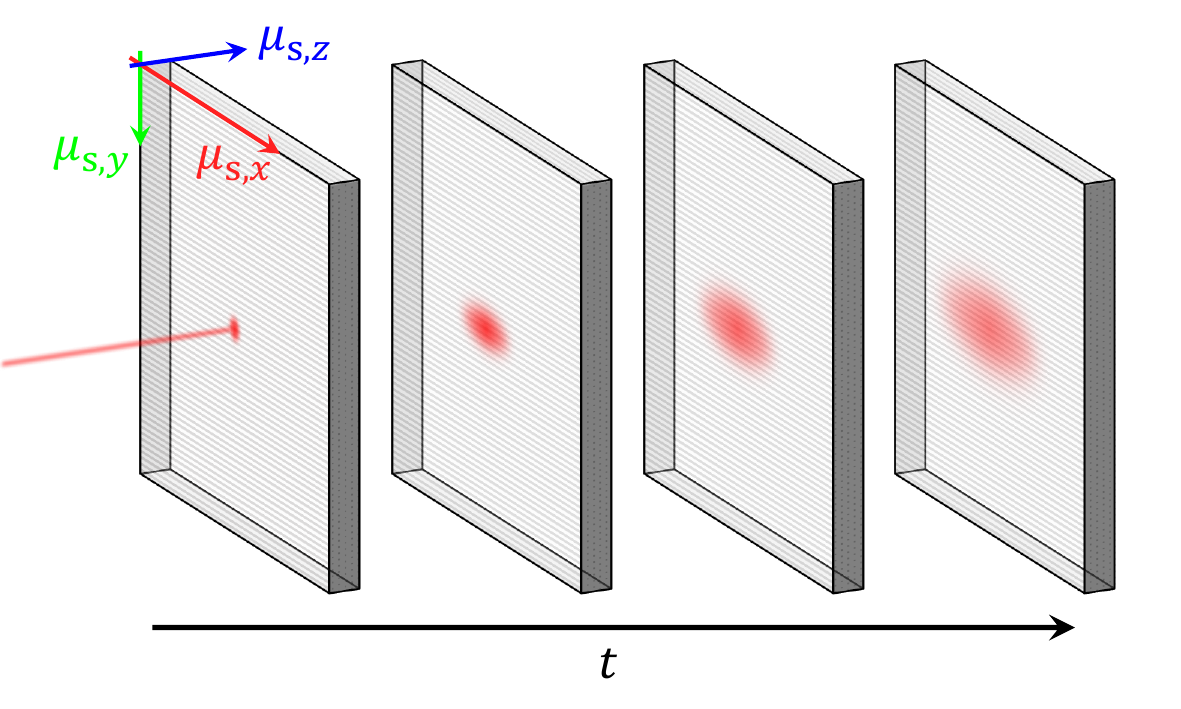}
	\caption{Schematic illustration of the time evolution of the diffuse intensity pattern at the surface of an anisotropic slab following excitation by a point-like pulsed source.
	Direction-dependent scattering coefficients ($\mux$, $\muy$, $\muz$) lead to anisotropic spreading of light, producing elongated diffusion profiles at increasing time delays.
	The coordinate system shown defines the reference frame used throughout this work.
	Adapted from \cite{pini2024experimental}.}
	\label{fig:anisPatt}
\end{figure}

Anisotropic light transport arises whenever the microscopic structure of a material exhibits directional order, leading to different scattering properties along distinct spatial directions and thus an elongated diffused pattern, as exemplified in Fig.~\ref{fig:anisPatt}.
Such behavior is observed across a wide range of systems, from biological tissues such as brain white matter \cite{pini2025anisotropic}, spinal cord \cite{depaoli2020anisotropic}, tendons \cite{simon2014time, nazarian2024structure}, teeth \cite{kienle2002light, zoller2018parallelized}, skin \cite{nickell2000anisotropy} and bone \cite{sviridov2005intensity}, to common materials such as wood \cite{kienle2008light, shanker2025spatiotemporally}, paper \cite{pini2024experimental} or compressed foam materials under mechanical deformations \cite{johnson2009optical, pini2024diffusion}.
Although anisotropic transport is a common feature in many diffusive media, its quantitative treatment remains elusive.
The fundamental difficulty lies in establishing a rigorous link between the microscopic scattering parameters and the macroscopic transport observables in the presence of structural anisotropy.
For this reason, even if anisotropic diffusion is detected, it is often disregarded or analyzed using oversimplified models.

An additional level of complexity is introduced when the single scattering phase function, i.e.\ the angular probability distribution after a scattering event, is also anisotropic.
To avoid confusion, we will refer to this anisotropy as single scattering ``asymmetry".
Similarly, the average cosine of the phase function $g=\expval{\cos\theta}$, which is sometimes known as the ``anisotropy factor", will be referred to as the ``asymmetry factor".
The phase function depends on the morphology and optical contrast of the scatterers and, for simple geometries such as spheres or cylinders, can be derived analytically from Mie theory.
For more complex systems, such as biological tissue, it is commonly approximated by the Henyey--Greenstein function.

In isotropic materials, light transport is well described by the classical diffusion equation, which links the scattering coefficient $\mus$ to the macroscopic diffusion constant $D$.
In the presence of scattering asymmetry $g\neq0$, the so-called similarity relation connects the scattering mean free path to its reduced counterpart $\musp=\mus(1-g)$, allowing the diffusion coefficient to be expressed in a compact analytic form $D=v/3\musp$.

However, in anisotropic media the diffusion process must be described by a tensor that couples the directional diffusion rates $\Dtens$ to the underlying scattering properties.
Common misconceptions in this case are represented by treating each diffusion coefficient as depending solely on the scattering coefficient along its corresponding direction, or by assuming that the similarity relation connecting scattering and reduced scattering coefficients is still applicable.

Several approaches have been proposed to describe anisotropic scattering beyond the diffusion approximation.
Among these, Mie theory applied to non-spherical or elongated scatterers provides a direct way to compute both the scattering coefficient and the angular phase function as functions of the incident direction with respect to the principal axis of the scatterer \cite{li2016gpu, zoller2018parallelized, liu2020polarized}.
While such microscopic models are physically accurate for highly dilute systems, their application to real materials is often impractical.
More frequently, one must introduce hybrid models comprising a mixture of spherical and cylindrical scatterers, specifying their relative concentrations, dimensions, and optical constants.
These assumptions introduce substantial complexity and degrees of freedom, making the inverse retrieval of optical parameters and the prediction of macroscopic observables cumbersome and highly model-dependent.

A more compact and versatile alternative that does not require any assumption on the sample microscopic structure is offered by the tensor-based description of anisotropic transport.
In this framework, the directional dependence of the scattering coefficient is represented by a tensor quantity $\mutens$, and the diffusion process is formulated as a generalized anisotropic diffusion equation (ADE).
This approach captures the essential features of structural anisotropy while retaining the simplicity and robustness of diffusion theory.
Importantly, it preserves analytical solvability, enabling closed-form solutions for both time-dependent and steady-state regimes.
This possibility has been previously exploited to derive analytical solutions for the special case of uniaxial anisotropy with symmetric scattering, which were validated both numerically and experimentally.

In this work, we extend the tensorial diffusion framework to the general case of full anisotropy and non-zero scattering asymmetry.
We consider this case specifically (with tensor scattering coefficients and scalar scattering asymmetry) as it represents a significant advancement over the previous description, while being still amenable to a fully analytical description which is likely not feasible when both $\mutens$ and $\gtens$ are independent tensors.
This generalization preserves the analytical form of the diffusion equation while accounting for more realistic microscopic features of scattering media.
Furthermore, the derived solutions include boundary conditions for bounded media (which also exhibit non-trivial dependence on $\mutens$ and $g$), accounting also for Fresnel reflections at the interfaces.
These expressions yield complete analytical formulas for evaluating time- and space-resolved transmission and reflection from the semi-infinite and slab geometries, offering a comprehensive framework for modeling light diffusion in structurally anisotropic systems.

\section{From Anisotropic Radiative Transfer to Diffusion Theory} \label{sec:ARTE}

At the microscopic level, structural anisotropy is expressed through a direction-dependent scattering coefficient $\mus(\vu{s})$, which depends on the unit vector $\vu{s}$ specifying the propagation direction of an energy packet prior to a scattering event, where $\vu{s} = (s_x,s_y,s_z) = (\sin\theta\cos\phi,\;\sin\theta\sin\phi,\;\cos\theta)$, with $\theta$ and $\phi$ as the polar and azimuthal angles in the absolute (laboratory) reference frame.
In the most general formulation of radiative transfer, other optical properties---including the absorption coefficient, refractive index, and single-scattering phase function---may also depend on direction.
While such a fully tensorial description is in principle admissible, it would introduce a large number of independent parameters and severe degeneracies, making both analytical treatment and experimental inference impractical.

The present work therefore focuses on a minimal yet physically expressive setting in which transport anisotropy is entirely encoded in the scattering process.
Specifically, the scattering coefficient is allowed to be direction dependent, while absorption, refractive index, and single-scattering asymmetry are treated as scalar parameters.
This choice retains the essential consequences of structural anisotropy for macroscopic transport while preserving analytical tractability and a clear connection to experimentally accessible parameters.

Under these assumptions, the extinction coefficient can be written as
\begin{equation}
	\muext(\vu{s}) = \mus(\vu{s}) + \mua,
\end{equation}
with $\mua$ denoting the absorption coefficient.
Reciprocity further implies
\begin{equation}
	\muext(-\vu{s}) = \muext(\vu{s}).
	\label{eq:reciprocity}
\end{equation}
Based on these definitions, we can define the Anisotropic Radiative Transfer Equation (ARTE) analogously to the RTE
\begin{equation}
	\frac{1}{v}\pdv{I(\vb*{r},t,\vu{s})}{t} =
	-\vu{s}\vdot \grad I(\vb*{r},t,\vu{s})
	-\muext(\vu{s})I(\vb*{r},t,\vu{s})
	+ \int_{4\piup} \mus(\vu{s}^\prime)\, I(\vb*{r},t,\vu{s}^\prime)\,
	p(\vu{s}^\prime,\vu{s}) \dd{\Omega^\prime}
	+ Q(\vb*{r},t,\vu{s}),
	\label{eq:ARTE}
\end{equation}
where $I(\vb*{r},t,\vu{s})$ is the specific intensity, $p(\vu{s}^\prime,\vu{s})$ the phase function, and $Q(\vb*{r},t,\vu{s})$ is the source term.

In the diffusive regime, the ARTE leads to a generalized anisotropic diffusion equation (ADE) \cite{vasques2014non, martelli2022light},
\begin{equation}
	\left(\pdv{t} - \div{\Dtens} \grad + v\mua\right)\Phi(\vb*{r},t) = v Q(\vb*{r},t),
\end{equation}
where $\Phi(\vb*{r},t)$ denotes the fluence rate and $\Dtens$ is the anisotropic diffusion tensor,
\begin{equation}
	\Dtens =
	\begin{pmatrix}
		D_{xx} & D_{xy} & D_{xz} \\
		D_{yx} & D_{yy} & D_{yz} \\
		D_{zx} & D_{zy} & D_{zz}
	\end{pmatrix}.
\end{equation}

In the following, we will assume that the directional dependence of the scattering coefficient can be expressed using nine scalar components of a scattering tensor $\mutens$,
\begin{equation}
	\mutens =
	\begin{pmatrix}
		\mu_{\mathrm{s},xx} & \mu_{\mathrm{s},xy} & \mu_{\mathrm{s},xz} \\
		\mu_{\mathrm{s},yx} & \mu_{\mathrm{s},yy} & \mu_{\mathrm{s},yz} \\
		\mu_{\mathrm{s},zx} & \mu_{\mathrm{s},zy} & \mu_{\mathrm{s},zz}
	\end{pmatrix},
\end{equation}
such that $\mus(\vu{s}) = \vu{s}\,\mutens\,\vu{s}^{\mathsf T}$.
While analogous tensorial descriptions could be introduced for absorption, refractive index, or single-scattering asymmetry, restricting anisotropy to the scattering coefficient alone already captures the dominant impact of structural organization on diffusive transport and leads to a mathematically closed and experimentally meaningful theory.

A general theoretical framework linking microscopic transport statistics to macroscopic diffusion has been developed by Vasques and Larsen \cite{vasques2014non}, encompassing both non-classical propagation regimes and angle-dependent step-length distributions. Within this formulation, anisotropic transport can be described through integral expressions for the diffusion tensor \cite{martelli2022light}.
However, explicit solutions to these expressions have not been reported nor validated against Monte Carlo simulations.

We first consider an isotropic phase function ($g=0$) to isolate the effect of direction-dependent scattering on the diffusion process.
The extension to the case with $g \neq 0$ is presented in Section~\ref{sec:asymmfact}.
Under these assumptions, the diffusion tensor elements can be written as
\begin{equation}
	\label{eq:Dij}
	D_{ij} = \frac{v}{4\piup \lavg}\int_{4\piup} \frac{s_i s_j}{\mus^2(\vu{s})} \dd{\Omega},
\end{equation}
where $i,j \in \{x,y,z\}$ and $\lavg$ denotes the direction-averaged mean free path.
Introducing $\ls(\vu{s}) = 1/\mus(\vu{s})$ as the direction-dependent scattering mean free path, one has
\begin{equation}
	\label{eq:ldef}
	\lavg = \int_{4\piup} \xi(\vu{s}) \ls(\vu{s}) \dd{\Omega},
\end{equation}
with $\xi(\vu{s})$ the probability density for a walker to travel within $\dd{\Omega}$ around direction $\vu{s}$ after a scattering event.
For an isotropic phase function that is independent of the incoming direction, and $\xi(\vu{s}) = 1/4\piup$.

If a set of principal axes of the microstructure can be identified and aligned with the laboratory Cartesian axes, the scattering tensor $\mutens$ may be taken as diagonal.
In this case, symmetry considerations in Eq.~\eqref{eq:Dij} imply that the off-diagonal diffusion coefficients vanish, and we adopt a more compact single-subscript notation
\begin{equation}
	\Dtens =
	\begin{pmatrix}
		D_x & 0 & 0 \\
		0 & D_y & 0 \\
		0 & 0 & D_z
	\end{pmatrix},
	\qquad
	\mutens =
	\begin{pmatrix}
		\mux & 0 & 0 \\
		0 & \muy & 0 \\
		0 & 0 & \muz
	\end{pmatrix}.
\end{equation}

In analogy with the isotropic diffusion relation, it is convenient to introduce direction-dependent reduced scattering coefficients $\muspi$ through
\begin{equation}
	\label{eq:muspdef}
	D_i = \frac{1}{3}\frac{v}{\muspi},
\end{equation}
and the associated transport mean free paths $\lt_i = 1/\muspi$.
It is important to stress that, even for $g = 0$, one generally has $\muspi \neq \mui$: in anisotropic media, no separable similarity relation exists that connects each $\mui$ to an independent ``reduced'' counterpart.
Accordingly, $\muspi$ (and $\lt_i$) should be regarded as a parametrization of the macroscopic diffusion tensor rather than as microscopic step-length parameters.
Neglecting this coupling between diffusion and the full set of scattering coefficients leads to a frequently used but generally incorrect application of the isotropic formula,
\begin{equation}
	\label{eq:Dsimpl}
	D_i^\text{simplistic} = \frac{1}{3}\frac{v}{\mui(1-g)}\neq D_i,
\end{equation}
which can introduce substantial errors in experimental inversion, with deviations that grow with increasing structural anisotropy.

\subsection{Connecting the microscopic properties to the macroscopic observables}

In a previous work \cite{pini2024diffusion}, we derived analytical solutions for the limited case of uniaxial anisotropy.
Here, we start by extending these solutions to the fully anisotropic scattering case, where the principal-axis coefficients can take values $\mux \neq \muy \neq \muz$, while the phase function remains isotropic ($g=0$).

The direction-dependent scattering coefficient can be written as
\begin{equation} \label{eq:dirdepmus}
	\mus(\vu{s})=\mus(\theta,\phi)=\mux\sin^2\theta\cos^2\phi+\muy\sin^2\theta\sin^2\phi+\muz\cos^2\theta.
\end{equation}

Using Eq.~\eqref{eq:ldef} with $\xi(\vu{s}) = 1/4\piup$ and $\ls(\vu{s}) = 1/\mus(\vu{s})$, the direction-averaged mean free path can be expressed in integral form as
\begin{equation} \label{eq:ldefint}
	\lavg = \frac{1}{4\piup} \int_0^{2\piup} \dd{\phi} \int_0^{\piup} \dd{\theta} \frac{\sin\theta}{\mus(\theta,\phi)} ,
\end{equation}
which admits the analytical solution
\begin{equation} \label{eq:lavg_anal}
	\lavg = \frac{F(\phi, m)}{2 \sqrt[4]{\mux \muy (\mux - \muz)(\muy - \muz)}},
\end{equation}
where $F(\phi,m)$ is the incomplete elliptic integral of the first kind and
\begin{equation} \label{eq:phim_def}
	\phi = 2\arctan \sqrt[4]{\frac{(\mux - \muz)(\muy - \muz)}{\mux \muy}}, \qquad
	m = \frac{1}{4} \left( 2 + \frac{2 \mux \muy - \muz \muy - \mux \muz}{\sqrt{\mux \muy (\mux - \muz)(\muy - \muz)}} \right).
\end{equation}

The diagonal diffusion coefficients follow from Eq.~\eqref{eq:Dij} by setting $i=j$:
\begin{align}
	D_x &= \frac{v}{4\piup\lavg} \int_{4\piup} \frac{s_x^2}{\mus^2(\vu{s})} \dd{\Omega} = \frac{v}{4\piup\lavg} \int_0^{2\piup} \dd{\phi} \int_0^{\piup} \dd{\theta} \frac{\sin^3\theta \cos^2\phi}{\mus^2(\theta, \phi)} \label{eq:Dx} \\
	D_y &= \frac{v}{4\piup\lavg} \int_{4\piup} \frac{s_y^2}{\mus^2(\vu{s})} \dd{\Omega} = \frac{v}{4\piup\lavg} \int_0^{2\piup} \dd{\phi} \int_0^{\piup} \dd{\theta} \frac{\sin^3\theta \sin^2\phi}{\mus^2(\theta, \phi)} \label{eq:Dy} \\
	D_z &= \frac{v}{4\piup\lavg} \int_{4\piup} \frac{s_z^2}{\mus^2(\vu{s})} \dd{\Omega} = \frac{v}{4\piup\lavg} \int_0^{2\piup} \dd{\phi} \int_0^{\piup} \dd{\theta} \frac{\sin\theta \cos^2\theta}{\mus^2(\theta, \phi)}. \label{eq:Dz}
\end{align}

These relations define the diffusion tensor for any set of principal-axis microscopic scattering coefficients.

From Eq.~\eqref{eq:dirdepmus}, one obtains directly
\begin{align}
	\sin^2\theta\cos^2\phi &= \pdv{}{\mux}\mus(\theta, \phi) \label{eq:dmux} \\
	\sin^2\theta\sin^2\phi &= \pdv{}{\muy}\mus(\theta, \phi) \label{eq:dmuy} \\
	\cos^2\theta &= \pdv{}{\muz}\mus(\theta, \phi). \label{eq:dmuz}
\end{align}

Using Eq.~\eqref{eq:dmux} together with the definition of $\lavg$ in Eq.~\eqref{eq:ldefint}, the integral expression for $D_x$ simplifies to
\begin{equation} \label{eq:dDx}
	D_x = \frac{v}{4\piup\lavg}\int_0^{2\piup} \dd{\phi} \int_0^{\piup} \dd{\theta} \frac{\sin^3\theta \cos^2\phi}{\mus^2(\theta, \phi)} = -\frac{v}{\lavg} \pdv{}{\mux} \lavg = -v\pdv{}{\mux} \ln\lavg.
\end{equation}

Similarly, Eqs.~\eqref{eq:dmuy} and \eqref{eq:dmuz} give
\begin{align}
	D_y &= -v\pdv{}{\muy} \ln\lavg \label{eq:Dy_log} \\
	D_z &= -v\pdv{}{\muz} \ln\lavg. \label{eq:Dz_log}
\end{align}

Substituting Eq.~\eqref{eq:lavg_anal} into Eqs.~\eqref{eq:dDx}, \eqref{eq:Dy_log}, or \eqref{eq:Dz_log} yields analytical expressions for the diffusion coefficients.
By symmetry under permutations of the principal axes, it is sufficient to report the result for a generic component $D_i$
\begin{equation}
	D_i = \frac{v}{2} \left[\frac{1}{\mui} \left( 1 - \frac{\sqrt{\muj \muk}}{\sqrt{(\muj - \mui)(\muk - \mui)}} \right) + \frac{1}{\lavg} \mathcal{A}_i \right]
	\label{Di_anal}
\end{equation}
where $(i,j,k)$ is any permutation of $(x,y,z)$ and
\begin{equation}\label{Bi_def}
	\mathcal{A}_i = \frac{1}{(\muj - \mui)(\muk - \mui) + \sqrt{\muj \muk (\muj - \mui)(\muk - \mui)}} + \frac{(\muj \muk)^{1/4}E(\phi_i,m_i)}{\mui \left[(\muj - \mui)(\muk - \mui)\right]^{3/4}}.
\end{equation}
with $E(\phi_i,m_i)$ denoting the incomplete elliptic integral of the second kind.
The arguments $\phi_i$ and $m_i$ follow from Eq.~\eqref{eq:phim_def} by substituting $(\mux,\muy,\muz) \rightarrow (\muj,\muk,\mui)$:
\begin{align}
	\phi_i &= \phi(\muj, \muk, \mui) \label{phimi_def} \\
	m_i &= m(\muj, \muk, \mui). \label{phimi_def_b}
\end{align}

Equation~\eqref{Di_anal} applies when $\mui\neq\muj\neq\muk$.
For parameter combinations where intermediate expressions become complex, the physically meaningful result is given by $D_i = \operatorname{Re}(D_i)$.
All expressions derived in this section reduce to the familiar isotropic limit as $\mux,\muy,\muz \rightarrow \mus$, as expected.

\begin{figure}
	\centering
	\includegraphics[width=0.5\linewidth]{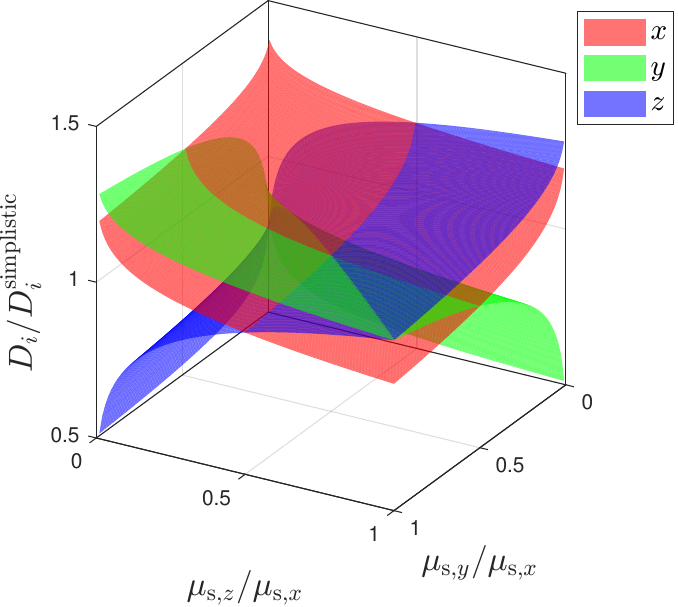}
	\caption{Ratios between the diffusion tensor elements and their corresponding simplistic counterparts for different values of $\muy/\mux$ and $\muz/\mux$, with $\mux$ fixed.
		This highlights the systematic bias introduced when anisotropy is ignored in diffusion-based parameter estimation.}
	\label{fig:DxDyDz}
\end{figure}

In Fig.~\ref{fig:DxDyDz} we report the ratio between the diffusion coefficients obtained from Eq.~\eqref{Di_anal} and their simplistic counterparts from Eq.~\eqref{eq:Dsimpl}, while varying $\muy/\mux$ and $\muz/\mux$ in $(0,2]$ at fixed $\mux$.
The isotropic case corresponds to the central point, where all curves converge to \num{1}.
The error introduced by using the simplistic formula Eq.~\eqref{eq:Dsimpl} increases arbitrarily with anisotropy and can become substantial in regimes relevant, e.g., to common fibrous materials.

\section{ADE solutions in the slab geometry for \texorpdfstring{\boldmath{$g=0$}}{g = 0}}

In the context of diffusion theory, a commonly studied configuration is that of a slab of thickness $L$, infinitely extended in the $xy$ plane, illuminated by a pencil beam pulse along the perpendicular $z$ axis, as illustrated in Fig.~\ref{fig:anisPatt}.
In the diffusive regime, a pencil beam source is more conveniently modeled as an isotropic point source $Q(\vb*{r}, t) = \delta(x) \delta(y) \delta(z - z_0) \delta(t)$ placed at the depth $z_0$ from the surface.
Using the method of virtual sources \cite{martelli2022light}, the solution of the time-dependent ADE can be written as:
\begin{equation} \label{eq:Urt}
	\Phi(\vb*{r},t)=
	\frac{\eu^{-\mua v t} \eu^{-x^2/(4D_x t)} \eu^{-y^2/(4D_y t)}}{(4\piup t)^{3/2}\sqrt{D_x D_y D_z}} 
	\sum_{m=-\infty}^{\infty} \left[ \eu^{-\frac{(z-z_{+,m})^2}{4D_z t}} - \eu^{-\frac{(z-z_{-,m})^2}{4D_z t}}\right]
\end{equation}
with
\begin{align*}
	z_{+, m} &= 2 m \left( L+2 \ze \right) + z_0, \\
	z_{-, m} &= 2 m \left( L+2 \ze \right) - 2 \ze - z_0,
\end{align*}
representing the position of the virtual sources, and $\ze$ as the extrapolated length.

\subsection{Boundary conditions}
In order to solve the ADE as a function of the microscopic scattering coefficient elements $\mui$, closed-form expressions for $z_0$ and extrapolated boundary length $\ze$ must be therefore derived.
	
\subsubsection{Equivalent isotropic source position \texorpdfstring{\boldmath{$z_0$}}{z0}}
The depth position of the equivalent isotropic point-like source $z_0$ is straightforward when assuming $g = 0$.
The depletion of a pencil beam ballistic component entering an anisotropic slab is determined solely by the scattering rate along $z$, yielding an exponential decay $\muz \exp(-z \muz)$, in direct analogy with the isotropic case $\mus \exp(-z \mus)$.
In this sense, $z_0$ represents the characteristic distance over which the transport process loses memory of the initial direction $\vu{z}$.
Since for $g = 0$ the phase function is uniform, this memory is lost after a single step, which gives
\begin{equation} \label{eq:z0}
	z_0 = \lz.
\end{equation}

\subsubsection{Extrapolated length \texorpdfstring{\boldmath{$\ze$}}{ze}}
The integral form for the extrapolated length $\ze$ must instead be derived using a general approach that accounts for an anisotropic radiance distribution and Fresnel reflection at the boundaries.

\begin{figure}
	\centering
	\includegraphics[width=\textwidth]{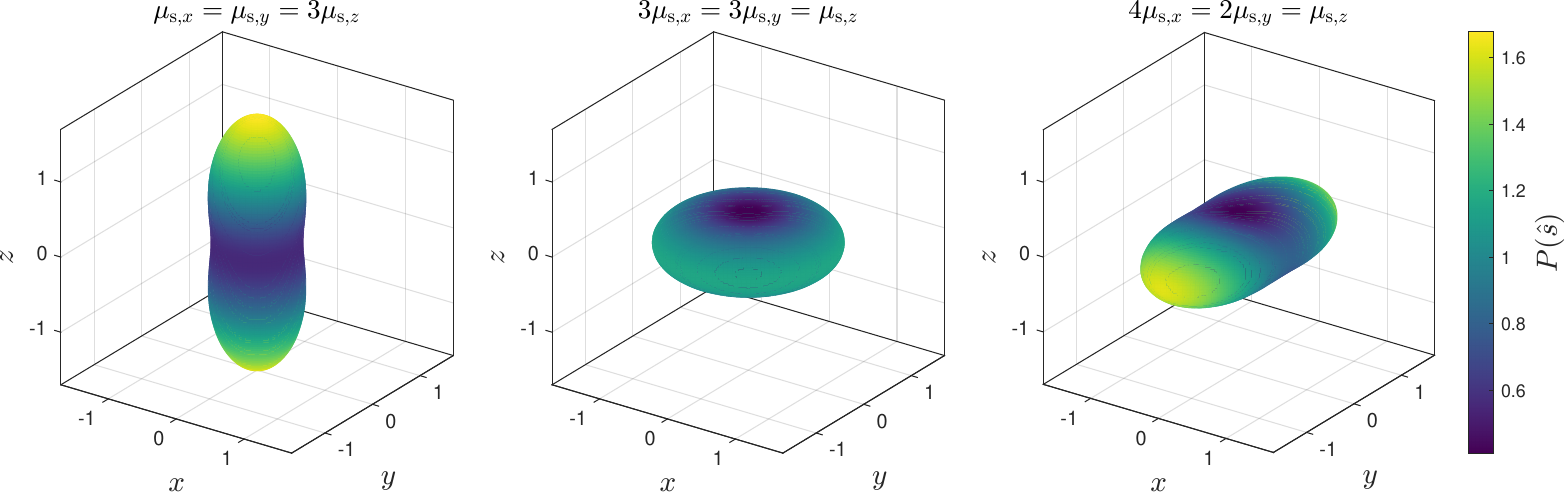}
	\caption{Representation of the angular distribution of the radiance $P(\vu{s})$ in the diffusive limit for different anisotropic scattering configurations.}
	\label{fig:anisPs}
\end{figure}

We first introduce the angular distribution of the radiance in the diffusive limit $P(\vu{s})$, defined as the probability density that a random walker propagates along $\vu{s}$ after many scattering events.
In the most general case, when both the step length and the phase function depend on direction, $P(\vu{s})$ has a non-trivial dependence on both \cite{alerstam2014anisotropic}.
For the present case, the diffusive-limit angular distribution depends only on $\mutens$ and reads
\begin{equation} \label{eq:Ps}
	P(\vu{s}) = \frac{1}{4\piup}\frac{\ls(\vu{s})}{\lavg} = \frac{1}{4\piup}\frac{1}{\lavg\mus(\vu{s})}.
\end{equation}
Notably, although Eq.~\eqref{eq:Ps} is introduced here for $g = 0$, it also holds for $g \neq 0$ under the present assumptions, because the phase function remains independent of the incoming direction.
Figure~\ref{fig:anisPs} shows examples of $P(\vu{s})$ for different illustrative configurations of structural anisotropy.

For a refractive index contrast $n = n_\text{in}/n_\text{ext}$ between the medium and the external environment, the polarization-averaged Fresnel reflectance for incident angle $\theta$ is
\begin{equation} \label{eq:Rfresnel}
	R(\theta) = \frac{1}{2}\left[\left(\frac{n\cos\theta - \sqrt{1 - n^2\sin^2\theta}}{n\cos\theta + \sqrt{1 - n^2\sin^2\theta}}\right)^2 + \left(\frac{n\sqrt{1 - n^2\sin^2\theta} - \cos\theta}{n\sqrt{1 - n^2\sin^2\theta} + \cos\theta}\right)^2\right].
\end{equation}
Following the notation of Alerstam \cite{alerstam2014anisotropic}, the extrapolated length can be expressed as
\begin{equation} \label{eq:ze}
	\ze = \frac{\Cze + \Yze}{\Bze - \Xze},
\end{equation}
where
\begin{align}
	\Bze &= \int_{\Omega_\text{up}} P(\vu{s}) s_z \dd{\Omega} = \frac{1}{4\piup\lavg}\int_0^{\piup/2} \dd{\theta} \int_0^{2\piup} \dd{\phi} \frac{\sin\theta \cos\theta}{\mus(\theta,\phi)}, \label{eqn:B} \\
	\Cze &= \int_{\Omega_\text{up}} P(\vu{s}) \ell(\vu{s}) s_z^2 \dd{\Omega} = \frac{1}{4\piup\lavg}\int_0^{\piup/2} \dd{\theta} \int_0^{2\piup} \dd{\phi} \frac{\sin\theta \cos^2\theta}{\mus^2(\theta,\phi)}, \label{eqn:C} \\
	\Xze &= \int_{\Omega_\text{up}} P(\vu{s}) s_z R(\theta) \dd{\Omega} = \frac{1}{4\piup\lavg}\int_0^{\piup/2} \dd{\theta} \int_0^{2\piup} \dd{\phi} \frac{\sin\theta \cos\theta R(\theta)}{\mus(\theta,\phi)}, \label{eq:X} \\
	\Yze &= \int_{\Omega_\text{up}} P(\vu{s}) \ls(\vu{s}) s_z^2 R(\theta) \dd{\Omega} = \frac{1}{4\piup\lavg}\int_0^{\piup/2} \dd{\theta} \int_0^{2\piup} \dd{\phi} \frac{\sin\theta \cos^2\theta R(\theta)}{\mus^2(\theta,\phi)}, \label{eq:Y}
\end{align}
where $\Omega_\text{up} = 2\piup$ denotes the solid angle of the upper hemisphere.

The analytical solution for $\Bze$ can be written as
\begin{equation} \label{eq:Banal}
	\Bze = -\frac{1}{4\piup\lavg}\frac{\acoth\left[\frac{2 \sqrt{\mux-\muz} \sqrt{\muy-\muz}}{\mux+\muy-2\muz}\right]+\acoth\left[\frac{2 \sqrt{\mux\muy(\mux-\muz)} \sqrt{\muy-\muz}}{-2\mux\muy+(\mux+\muy)\muz}\right]}{4 \sqrt{\mux-\muz} \sqrt{\muy-\muz}}.
\end{equation}
For $\Cze$ it can be shown that
\begin{equation} \label{eq:CfromDz}
	\Cze = \frac{D_z}{2v}.
\end{equation}
The integrals $\Xze$ and $\Yze$, on the other hand, must be typically evaluated numerically, while they vanish for a matched refractive index ($n = 1$).

Equivalently, the extrapolated length can be expressed in the unified integral form
\begin{equation} \label{eq:ze3D}
	\ze = \frac{\displaystyle\int_0^{\piup/2} \dd{\theta} \int_0^{2\piup} \dd{\phi} \frac{\sin\theta \cos^2\theta}{\mus^2(\theta,\phi)} [1 + R(\theta)]}{\displaystyle\int_0^{\piup/2} \dd{\theta} \int_0^{2\piup} \dd{\phi} \frac{\sin\theta \cos\theta}{\mus(\theta,\phi)} [1 - R(\theta)]}.
\end{equation}
	
\section{ADE solutions in the slab geometry for \texorpdfstring{\boldmath{$g \neq 0$}}{g ≠ 0}} \label{sec:asymmfact}

We now extend the anisotropic diffusion framework to include the effect of a scalar single-scattering asymmetry factor $g$.
Although in structurally anisotropic systems the phase function may depend on direction, leading to a tensorial asymmetry factor, we deliberately restrict our description to a scalar $g$.
This choice avoids cross-talk between multiple tensorial quantities, which would generally prevent the unique determination of optical parameters, reflects the absence of a consolidated framework for direction-dependent asymmetry factors in the literature, and simplifies the problem by preserving the azimuthal symmetry of the phase function.

A widely adopted model for single scattering in complex media is the Henyey--Greenstein phase function,
\begin{equation} \label{eq:HGF}
	p(\cos\theta) = \frac{1 - g^2}{4\piup(1 + g^2 - 2g\cos\theta)^{3/2}},
\end{equation}
which has proven effective in describing multiple scattering in many realistic systems, including biological tissues \cite{martelli2022light}.

Because the phase function remains independent of the incoming direction prior to scattering, the direction-averaged mean free path $\expval{\ell}$ retains the same form as in the $g=0$ case (Eq.~\eqref{eq:lavg_anal}).
For the same reason, the angular distribution of the radiance $P(\vu{s})$ remains unchanged (Eq.~\eqref{eq:Ps}).
Conversely, the diffusion tensor must be re-derived, because scattering asymmetry introduces persistent angular correlations that affect the macroscopic transport rates.
The full derivation is presented in Appendix~\ref{appA}.
Here we report the final result for a generic diagonal diffusion tensor element $D_i$, expressed as an infinite expansion over spherical harmonics $Y_l^m(\vu{s})$
\begin{equation} \label{eq:Di_g}
	D_i = \eval{D_i}_{g = 0} + \frac{v}{4\piup \lavg}\sum_{n = 0}^\infty \frac{g^{2n + 1}}{1 - g^{2n + 1}} \sum_{m=-n}^n \abs{H^i_{2n+1, 2m}}^2,
\end{equation}
where
\begin{equation} \label{eq:Hilm_g}
	H^i_{lm} = \int_{4\piup} \frac{s_i^\prime Y_l^m(\vu{s}^\prime)}{\mus(\vu{s}^\prime)} \dd{\Omega^\prime}.
\end{equation}

\begin{figure}
	\centering
	\begin{subfigure}{0.48\textwidth}
		\centering
		\includegraphics[width=\textwidth]{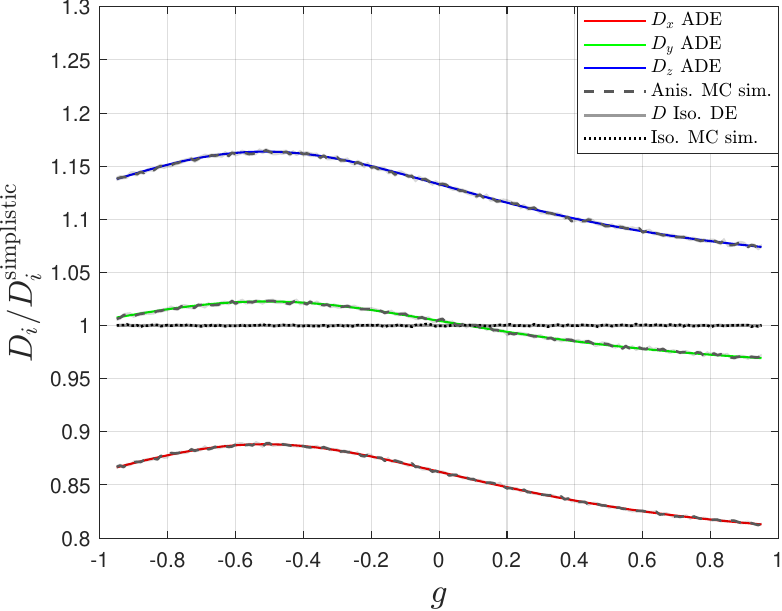}
	\end{subfigure}
	\hfill
	\begin{subfigure}{0.48\textwidth}
		\centering
		\includegraphics[width=\textwidth]{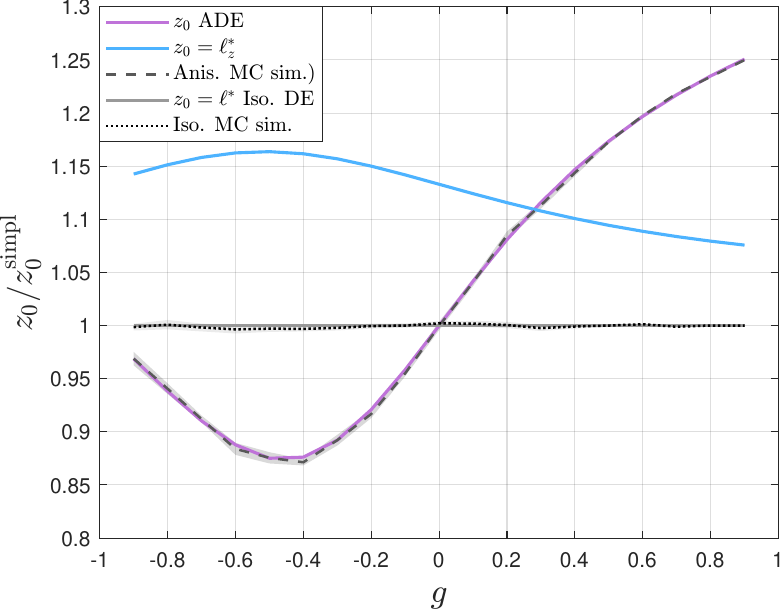}
	\end{subfigure}
	\vspace{2mm}
	\caption{(a) Ratios between the diffusion tensor elements and their corresponding simplistic approximations as a function of the asymmetry factor $g$.
	Solid colored curves denote the analytical predictions, while black dashed curves show anisotropic Monte Carlo results.
	Shaded areas indicate $2\sigma$ confidence intervals.
	Results are shown for $4\mux=2\muy=\muz$, no free parameters are used in the analytical curves.
	The gray curve represents the isotropic case.
	(b) Corresponding ratios for the equivalent isotropic source position $z_0$.
	The light blue curve representing the previously proposed approximation \cite{alerstam2014anisotropic} $z_0=\lz^\ast$ is shown for comparison.}
	\label{fig:ADEgvsMC}
\end{figure}

These expressions were validated against Monte Carlo simulations of random walkers propagating in an infinite anisotropic scattering medium over a wide range of asymmetry factors $g$.
Figure~\ref{fig:ADEgvsMC}a shows the ratio between the diffusion tensor elements and the commonly used simplistic expressions.
For the illustrative anisotropic configuration $4\mux = 2\muy = \muz$, at least one diffusion coefficient deviates by more than \qty{10}{\percent} across the entire range of $-1 < g < 1$.
By contrast, in the isotropic limit $\mus(\vu{s}) = \mus$, the similarity relation remains exact for all values of $g$, in full agreement with Monte Carlo results.
In this case, Eq.~\eqref{eq:Di_g} correctly reduces to the standard expression $D = v/[3\mus(1-g)]$.

\subsection{Boundary conditions}
When scattering asymmetry is present, both the equivalent isotropic source depth $z_0$ and the extrapolated length $\ze$ acquire a non-trivial dependence on $g$ and on the full scattering tensor.
For $g \neq 0$, the microscopic mean free path $\ell(\vu{s})$ entering the boundary-condition integrals $\Bze$, $\Cze$, $\Xze$ and $\Yze$ (Eqs.~\eqref{eqn:B}, \eqref{eqn:C}, \eqref{eq:X} and \eqref{eq:Y}) no longer represents the correct length scale.
A previously proposed approximation to overcome the simplistic approach replaces $\ell(\vu{s})$ with the transport length $\lt(\vu{s})$ inferred from the diffusion tensor and assumes $z_0 = \lz^\ast$ \cite{alerstam2014anisotropic}.
However, this approach neglects the finite angular persistence induced by the phase function and becomes increasingly inaccurate for small $g$ and strong anisotropy.

\subsubsection{Equivalent isotropic source position \texorpdfstring{\boldmath{$z_0$}}{z0}}
For $g \neq 0$, the source depth $z_0$ is linked to the cumulative directional memory of the random walk rather than its asymptotic diffusion rate.
A new length scale must therefore be evaluated explicitly from the underlying scattering kernel.
To this end, we introduce a \emph{direction-dependent persistence length} $\lambda(\vu{s})$, representing the characteristic distance over which the transport process loses memory of the initial direction $\vu{s}$.
The full derivation is given in Appendix~\ref{appB}.
For a Henyey--Greenstein phase function, $\lambda(\vu{s})$ can be written as
\begin{equation}
	\label{eq:perslen}
	\lambda(\vu{s})=\sum_{n=0}^{\infty}\frac{4n+3}{4\piup} \frac{1}{1-g^{2n+1}}\int_{4\piup}\frac{\vu{s}\cdot\vu{s}^\prime P_{2n+1}(\vu{s}\cdot\vu{s}^\prime)}{\mus(\vu{s}^\prime)} \dd{\Omega^\prime},
\end{equation}
where $P_{2n+1}$ denotes the Legendre polynomial of order $2n+1$.
The equivalent isotropic source depth then generalizes to
\begin{equation}
	\label{eq:z0_g}
	z_0 = \lambda_z = \lambda(\vu{z}).
\end{equation}

The accuracy of Eq.~\eqref{eq:z0_g} was assessed by Monte Carlo simulations in which random walkers are initialized with a pencil-beam angular distribution along $\vu{z}$ and propagated in an infinite anisotropic scattering medium.
After a sufficiently large number of scattering events, the ensemble-averaged displacement along $z$ converges to $z_0$.
Results are shown in Fig.~\ref{fig:ADEgvsMC}b as a ratio with respect to the commonly used approximation $z_0^\text{simplistic} = \ell_z/(1-g)$.

\subsubsection{Extrapolated length \texorpdfstring{\boldmath{$\ze$}}{ze}}
The extrapolated length $\ze$ depends on the asymptotic diffusion rate along $z$ and on the refractive index contrast $n$.
Since the stationary radiance distribution $P(\vu{s})$ remains independent of $g$ under the present assumptions, the boundary integrals $\Bze$ and $\Xze$ are unchanged.
The integral $\Cze$ is directly linked to $D_z$ via Eq.~\eqref{eq:CfromDz}, and therefore the diffusion rate along $z$ must be evaluated using Eq.~\eqref{eq:Di_g}.
The Fresnel-weighted term $\Yze$ accounts for boundary reflections, and its general expression becomes
\begin{equation} \label{eq:Y_final}
	\Yze = \eval{\Yze}_{g = 0} + \frac{1}{4\piup\lavg}\sum_{n=0}^{\infty}\frac{g^{2n+1}}{1-g^{2n+1}}\sum_{m=-n}^n H_{2n+1,m}^z \widetilde H_{2n+1,m}^z,
\end{equation}
where $H_{2n+1,m}^z$ denotes the $z$ component of Eq.~\eqref{eq:Hilm_g}, and
\begin{equation}\label{eq:Htilde_R_n}
	\widetilde{ H}^{z}_{2n+1,m} = \int_{\Omega_\text{up}} \frac{s_z}{\mus(\vu{s})} Y_{2n+1,m}^\ast (\vu{s}) R(\theta) \dd{\Omega}
\end{equation}
with $R(\theta)$ as the Fresnel reflectance defined in Eq.~\eqref{eq:Rfresnel}.
The term $\eval{\Yze}_{g = 0}$ corresponds to the isotropic-scattering contribution and includes the same Fresnel weighting over the upper hemisphere.
A detailed derivation of Eq.~\eqref{eq:Y_final}, including the hemisphere-restricted angular projections and the inclusion of Fresnel reflections, is provided in Appendix~\ref{appC}.
Finally, the extrapolated length is evaluated using Eq.~\eqref{eq:ze}.
%Since $\ze$ has no direct counterpart among the input parameters of a Monte Carlo simulation, the validity of its expression is assessed through the agreement between the ADE predictions and the Monte Carlo results in bounded geometries.

\section{Time- and space-dependent solutions of ADE} \label{sec:anisMC}

We validate the results presented in the previous section using the open source Monte Carlo package PyXOpto \cite{naglic2021pyxopto}, which has been modified to handle anisotropic scattering and asymmetry factors.
The ADE solutions presented here in the time and space domains are also made available as a set of Python and MATLAB functions at \url{https://github.com/epini/ADE}, where the integrals to evaluate the macroscopic transport properties are solved numerically.

Time-resolved reflected and transmitted intensity profiles for a slab illuminated along the $z$ axis can be obtained from the fluence rate in Eq.~\eqref{eq:Urt} using Fick's law \cite{haskell1994boundary, martelli2022light} by setting
\begin{align}
	R(x, y, t) &= \frac{D_z}{v} \pdv{z} \Phi(x, y, z=0, t) \label{eq:Rrt}, \\
	T(x, y, t) &= -\frac{D_z}{v} \pdv{z} \Phi(x, y, z=L, t) \label{eq:Trt},
\end{align}
resulting in bi-variate Gaussian distributions:
\begin{multline} \label{eq:anisrefl}
	R(x,y,t)=-\frac{v\exp(-\mua vt) \exp(-\frac{x^2}{4D_xt}) \exp(-\frac{y^2}{4D_yt})}{2(4\piup)^{3/2}t^{5/2} \sqrt{D_x D_y D_z}}\times \\
	\times\sum_{m=-\infty}^{\infty} \left[z_{3,m} \exp(-\frac{z_{3,m}^2}{4D_zt} )-z_{4,m} \exp(-\frac{z_{4,m}^2}{4D_zt} )\right],
\end{multline}
\begin{multline} \label{eq:anistransm}
	T(x,y,t) = \frac{v\exp(-\mua vt) \exp(-\frac{x^2}{4D_xt}) \exp(-\frac{y^2}{4D_yt})}{2(4\piup)^{3/2} t^{5/2} \sqrt{D_x D_y D_z}} \times \\
	\times \sum_{m=-\infty}^{\infty} \left[z_{1,m} \exp(-\frac{z_{1,m}^2}{4D_zt})-z_{2,m} \exp(-\frac{z_{2,m}^2}{4D_zt})\right],
\end{multline}
where
\begin{align*}
	z_{1,m} &= L(1-2m) - 4m\ze - z_0,\\
	z_{2,m} &= L(1-2m) - (4m-2)\ze + z_0,\\
	z_{3,m} &= -2mL - 4m\ze - z_0,\\
	z_{4,m} &= -2mL - 4(m-2)\ze + z_0.
\end{align*}

The Mean Square Displacements (MSD) along $x$ and $y$ are
\begin{equation} \label{eq:MSDdef}
	w_x^2 (t) = 4 D_x t, \qquad w_y^2(t) = 4 D_y t.
\end{equation}

Accessing either $R(x, y, t)$ or $T(x, y, t)$ at different times enables the direct retrieval of the diffusion rates along $x$ and $y$.
Moreover, the evolution of $w_x^2(t)$ and $w_y^2(t)$ does not depend on the amplitude of the profile, so that any effect that may modify its overall intensity, such as a homogeneous absorption coefficient, factors out exactly from the MSD.

Total time-resolved reflectance and transmittance expressions can be derived by direct integration of Eq.~\eqref{eq:Trt} in space
\begin{equation} \label{eq:anisrefl_t}
	R(t) = \frac{\exp(-\mua vt)}{2(4\piup D_z)^{1/2}t^{3/2}}\sum_{m=-\infty}^{\infty} \left[z_{3,m} \exp(-\frac{z_{3,m}^2}{4D_zt})-z_{4,m} \exp(-\frac{z_{4,m}^2}{4D_zt}) \right],
\end{equation}
\begin{equation} \label{eq:anistransm_t}
	T(t) = \frac{\exp(-\mua vt)}{2(4\piup D_z)^{1/2}t^{3/2}} \sum_{m=-\infty}^{\infty} \left[z_{1,m}\exp(-\frac{z_{1,m}^2}{4D_zt})-z_{2,m} \exp(-\frac{z_{2,m}^2}{4D_zt}) \right].
\end{equation}
The steady state profiles are obtained integrating Eq.~\eqref{eq:Trt} in time
\begin{multline} \label{eq:anissteadystateR}
	R(x,y) = -\frac{1}{4\piup\sqrt{D_x D_y D_z}} \sum_{m=-\infty}^{\infty} \Bigg[ z_{3,m} \left(\frac{x^2}{D_x} + \frac{y^2}{D_y} + \frac{z_{3,m}^2}{D_z}\right)^{-3/2} \left(1 + \sqrt{\mua v\left(\frac{x^2}{D_x} + \frac{y^2}{D_y} + \frac{z_{3,m}^2}{D_z}\right)}\right) \times \\
	\times \exp(-\sqrt{\mua v \left(\frac{x^2}{D_x} + \frac{y^2}{D_y} + \frac{z_{3,m}^2}{D_z}\right)})
	-z_{4,m} \left(\frac{x^2}{D_x} + \frac{y^2}{D_y} + \frac{z_{4,m}^2}{D_z}\right)^{-3/2} \times \\
	\times \left(1 + \sqrt{\mua v \left(\frac{x^2}{D_x} + \frac{y^2}{D_y} + \frac{z_{4,m}^2}{D_z}\right)}\right) \exp(-\sqrt{\mua v\left(\frac{x^2}{D_x} + \frac{y^2}{D_y} + \frac{z_{4,m}^2}{D_z}\right)}) \Bigg],
\end{multline}
\begin{multline} \label{eq:anissteadystateT}
	T(x,y) = \frac{1}{4\piup\sqrt{D_x D_y D_z}} \sum_{m=-\infty}^{\infty} \Bigg[ z_{1,m} \left( \frac{x^2}{D_x} + \frac{y^2}{D_y} + \frac{z_{1,m}^2}{D_z} \right)^{-3/2} \left(1+\sqrt{\mua v \left( \frac{x^2}{D_x} + \frac{y^2}{D_y}+\frac{z_{1,m}^2}{D_z} \right)}\right)\times \\
	\times \exp(-\sqrt{\mua v \left(\frac{x^2}{D_x} + \frac{y^2}{D_y} + \frac{z_{1,m}^2}{D_z}\right)})
	-z_{2,m} \left(\frac{x^2}{D_x} + \frac{y^2}{D_y} + \frac{z_{2,m}^2}{D_z}\right)^{-3/2}\times \\
	\times \left(1 + \sqrt{\mua v \left(\frac{x^2}{D_x} + \frac{y^2}{D_y} + \frac{z_{2,m}^2}{D_z}\right)}\right) \exp(-\sqrt{\mua v \left(\frac{x^2}{D_x} + \frac{y^2}{D_y} + \frac{z_{2,m}^2}{D_z}\right)}) \Bigg].
\end{multline}
Finally, integrating in time Eq.~\eqref{eq:anisrefl_t} and \eqref{eq:anistransm_t} gives the total reflectance and transmittance
\begin{equation} \label{eq:anisTR}
	R = -\frac{1}{2} \sum_{m=-\infty}^{\infty} \left[\sgn (z_{3,m}) \exp(-\abs{z_{3,m}} \sqrt{\frac{\mua v}{D_z}}) - \sgn (z_{4,m}) \exp(-\abs{z_{4,m}} \sqrt{\frac{\mua v}{D_z}})\right],
\end{equation}
\begin{equation} \label{eq:anisTT}
	T = \frac{1}{2} \sum_{m=-\infty}^{\infty} \left[\sgn (z_{1,m}) \exp(-\abs{z_{1,m}} \sqrt{\frac{\mua v}{D_z}}) - \sgn (z_{2,m}) \exp(-\abs{z_{2,m}} \sqrt{\frac{\mua v}{D_z}})\right].
\end{equation}
Equations~\eqref{eq:anisTR} and \eqref{eq:anisTT} can be used only for $\mua > 0$, thus the total absorbance can be evaluated as $A=1-R-T$.
In the case of a non-absorbing optically thick slab ($L \gg \lavg$) the usual scaling of the total transmittance with thickness holds as
\begin{equation} \label{eq:totalanistransmission}
	T = \frac{z_0 + \ze}{L + 2\ze}.
\end{equation}
Equation~\eqref{eq:totalanistransmission} is analogous to the isotropic formula for total transmission \cite{martelli2022light}, where the presence of anisotropy is now taken into account in $\ze$ and $z_0$.
Therefore, the total reflectance can be retrieved as $R=1-T$.

\begin{figure}
	\centering
	\includegraphics[width=0.8\textwidth]{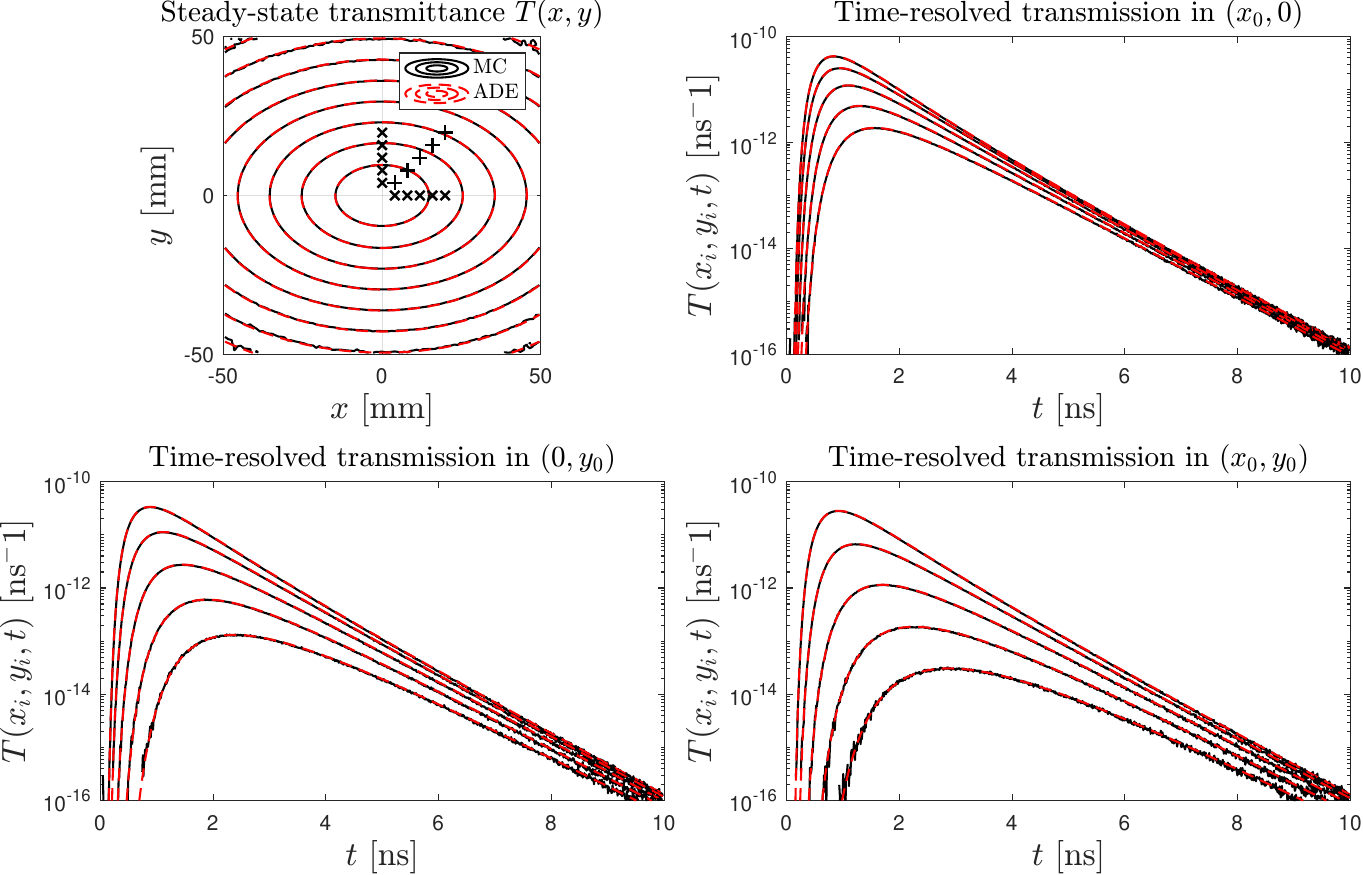}
	\caption{(a) Steady-state transmittance $T(x, y)$ from a \qty{10}{\milli\meter}-thick slab with $\mux = \qty{2.5}{\per\milli\meter}$, $\muy = \qty{7.5}{\per\milli\meter}$, $\muz = \qty{10}{\per\milli\meter}$, $n = 1.3$, $\mua=\qty{2e-6}{\per\milli\meter}$, and $g=0$.
	Data are plotted as log-scale contour lines marking consecutive decades.
	(b, c, d) Time-resolved transmittance at different locations $(x_0, y_0)$ on the slab's output surface, identified by the markers in panel (a).
	The prediction from anisotropic theory (dashed) is compared against the results of Monte Carlo simulations with \num{e12} trajectories.}
	\label{fig:TSres}
\end{figure}

\begin{figure}
	\centering
	\includegraphics[width=0.8\textwidth]{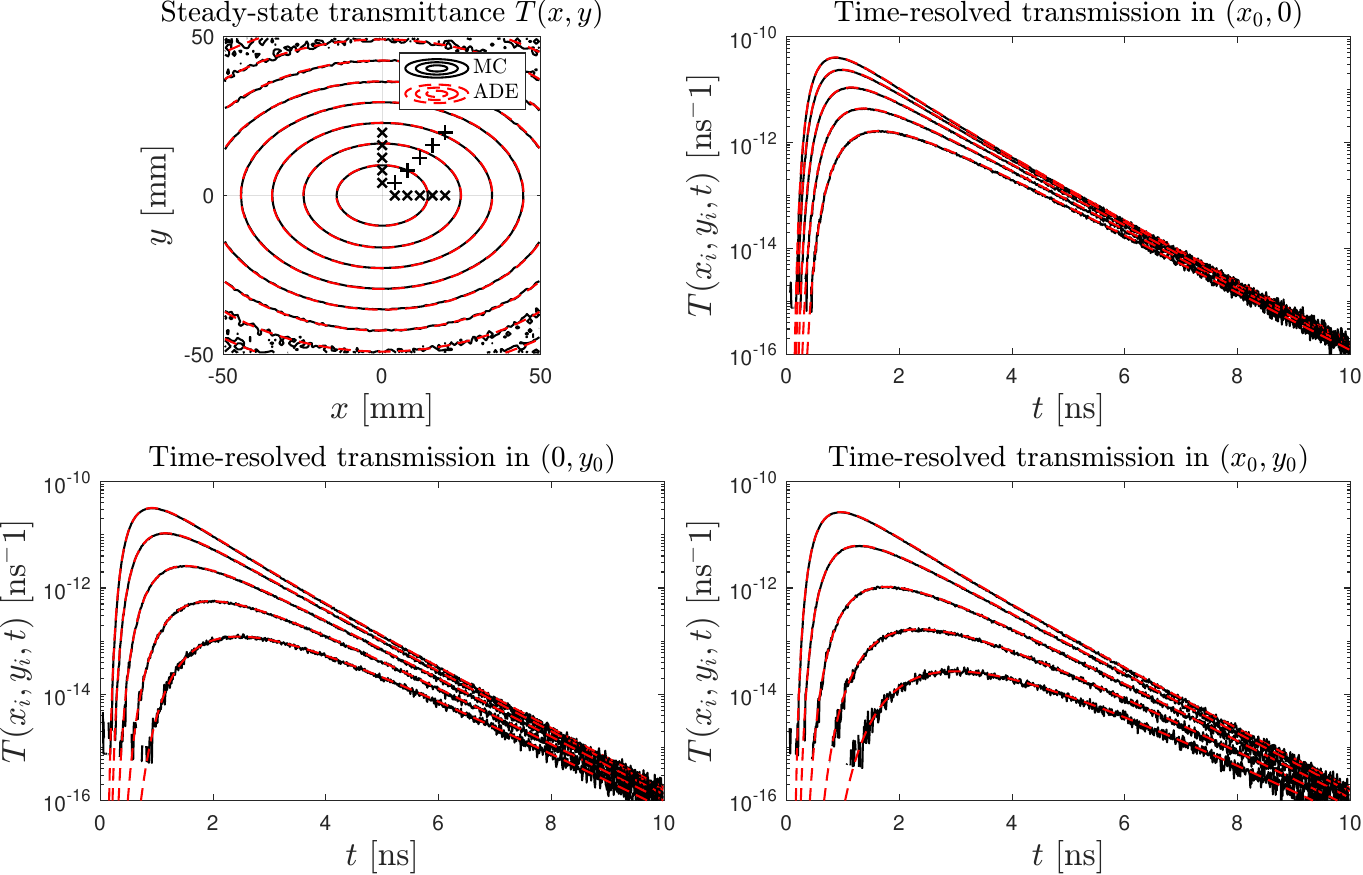}
	\caption{(a) Steady-state transmittance $T(x, y)$ from a \qty{10}{\milli\meter}-thick slab with $\mux = \qty{25}{\per\milli\meter}$, $\muy = \qty{75}{\per\milli\meter}$, $\muz = \qty{100}{\per\milli\meter}$, $n = 1.3$, $\mua=\qty{2e-6}{\per\milli\meter}$, and $g=0.9$.
	Data are plotted as log-scale contour lines marking consecutive decades.
	(b, c, d) Time-resolved transmittance at different locations $(x_0, y_0)$ on the slab's output surface, identified by the markers in panel (a).
	The prediction from anisotropic theory (dashed) is compared against the results of Monte Carlo simulations.
	Due to the longer computational effort required by high-$g$ values, only \num{e11} trajectories have been simulated in this case.}
	\label{fig:TSres09}
\end{figure}

To validate these solutions, we ran a set of MC simulations for the time- and space-resolved transmittance $T(x, y, t)$ for anisotropic slabs with various optical properties.
We report the results for two representative cases with full anisotropy, mismatched refractive index contrast, non-zero absorption, and $g=\{\num{0}, \num{0.9}\}$.
The Monte Carlo outputs are compared to the numerically evaluated ADE predictions (Figures \ref{fig:TSres} and \ref{fig:TSres09}), showing excellent agreement in both cases.
%Since for $g=0$ the solution is exact, excellent agreement is found at all times and positions, with average relative discrepancies below \qty{1}{\percent} in the considered spatio-temporal range.
%In Fig.~\ref{fig:TSres09}a-d) we report results for a similar simulation with $g=0.9$, in this case the analytical solution is approximated, however the agreement with the simulation is very good for all curves.

\section{Discussion}
In this work, we have developed a generalized diffusion framework for light transport in fully anisotropic scattering media.
By linking a tensorial description of the microscopic scattering coefficient to macroscopic diffusion rates, we have shown that anisotropic transport cannot be described by applying the isotropic diffusion formula independently along each axis.
A central result is that each diagonal element of the diffusion tensor depends on all components of the scattering tensor.
This coupling becomes increasingly relevant with growing anisotropy and can lead to substantial errors when simplistic substitutions are used.
The analytical relations derived here, validated against anisotropic Monte Carlo simulations, provide a compact and rigorous framework that retains the form and robustness of diffusion-based modeling.

The inclusion of a scalar single-scattering asymmetry factor extends the model to realistic systems while preserving analytical tractability.
Although closed-form expressions are not available for $g \neq 0$, the corresponding analytical series converges rapidly and yields diffusion coefficients in excellent agreement with numerical simulations across the full range of scattering asymmetry $-1 < g < 1$.

Additionally, we have shown that in bounded anisotropic system, $z_0$ and $\ze$ depend explicitly on microscopic scattering properties rather than solely on transport mean free paths, highlighting the inapplicability of boundary conditions inherited from isotropic theory.
In the presence of scattering asymmetry, $z_0$ is controlled by a direction-dependent persistence length that quantifies the cumulative angular memory of the random walk, whereas $\ze$ is governed by the asymptotic diffusion rate along the boundary normal together with the refractive index contrast.
These results provide a physically consistent interpretation of anisotropic boundary conditions and enable diffusion solutions that accurately reproduce Monte Carlo predictions for time- and space-resolved reflectance and transmittance.
From an experimental perspective, these closed-form slab solutions offer a direct route to retrieving anisotropic transport parameters from measured data.

Finally, although the present framework is formulated for radiative transport, its structure is not specific to optics.
Closely related transport and diffusion equations arise in the propagation of neutrons and other ionizing particles \cite{vasques2014non}, as well as in diffusive heat transport in anisotropic media \cite{nouri2015three, wang2025three}, where direction-dependent cross sections or mean free paths give rise to tensorial diffusion coefficients.
In this broader context, the analytical relations derived here may help connect microscopic transport properties to macroscopic diffusion behavior beyond commonly adopted approximations.

\medskip

% Acknowledgements
\medskip
\textbf{Acknowledgements} \par %delete if not applicable))
The authors further thank Fabrizio Martelli, Andre Liemert and Alwin Kienle for fruitful discussion.
This work was partially funded by the European European Union's NextGenerationEU Programme with the I-PHOQS Research Infrastructure [IR0000016, ID D2B8D520, CUP B53C22001750006] ``Integrated infrastructure initiative in Photonic and Quantum Sciences''.
L.P.\ acknowledges the CINECA award under the ISCRA initiative, for the availability of high performance computing resources and support (ISCRA-C ``ARTTESC2'').

% Acknowledgements
\medskip
\textbf{Conflict of Interest} \par %delete if not applicable))
The authors declare no conflict of interest.

% Acknowledgements
\medskip
\textbf{Data Availability Statement} \par %delete if not applicable))
The scripts and functions that support the findings of this study are openly available on \url{https://github.com/epini/ADE}.
Anisotropic Monte Carlo simulations have been performed using a open-source software tool \url{https://github.com/xopto/pyxopto}.

% References
\medskip

% Use the following code if you wish to generate your bibliography with BibTeX;
% replace the string "MSP-template" below with the name(s) of
% the BibTeX data base(s) you want to use.
% The resulting bibliography-output (the content of the .bbl file)
% must be pasted back into this file before submission.
% Please also include your BibTeX data base file(s) in your submission
% so that we can re-run BibTeX if necessary.
%
\bibliographystyle{MSP}
\bibliography{main}

\appendix
\section{Diffusion tensor components for a scalar asymmetry factor} \label{appA}
For the case of anisotropic diffusion with $g \neq 0$, the diffusion tensor components can be written as \cite{martelli2022light}
\begin{equation} \label{eq:Dijgneq0}
	D_{ij} = \frac{v}{4\piup\lavg} \int_{4\piup} \left( \frac{s_i}{\mus^2(\vu{s})} - \frac{\zeta^i(\vu{s})}{\mus(\vu{s})}\right) s_j \dd{\Omega},
\end{equation}
where
\begin{equation}\label{eq:zeta}
	\zeta^i (\vu{s}) = \sum_{n = 0}^\infty \zeta_n^i (\vu{s}),
\end{equation}
with
\begin{equation} \label{eq:zeta0}
	\zeta^i_0(\vu{s}) = \mathcal{S}^i(\vu{s}) = - \int_{4\piup} \dd{\Omega^\prime} s^\prime_i \frac{p(\vu{s}\cdot \vu{s}^\prime)}{\mus(\vu{s}^\prime)} ,
\end{equation}
and
\begin{equation} \label{eq:zetan}
	\zeta^i_{n+1} (\vu{s}) = \int_{4\piup} \dd{\Omega^\prime} p(\vu{s}\cdot \vu{s}^\prime) \zeta_n^i(\vu{s}^\prime).
\end{equation}

The Henyey--Greenstein phase function $p(\vu{s}\cdot \vu{s}^\prime)$ can be written in terms of spherical harmonics as
\begin{equation} \label{eq:pHG}
	p(\vu{s}\cdot \vu{s}^\prime) = \sum_{l= 0}^\infty\frac{(2l + 1)}{4\piup} g^l P_l(\vu{s}\cdot \vu{s}^\prime) = \sum_{l= 0}^\infty \sum_{m = -l}^l g^l Y_l^{m \ast}(\vu{s})Y_l^m(\vu{s}^\prime),
\end{equation}
where we have used the addition theorem of spherical harmonics
\begin{equation}
	\frac{(2l + 1)}{4\piup} P_l (\vu{s}\cdot \vu{s}^\prime) = \sum_{m = -l}^l Y_l^{m \ast}(\vu{s})Y_l^m(\vu{s}^\prime).
\end{equation}

By substituting Eq.~\eqref{eq:pHG} in Eq.~\eqref{eq:zeta0} we find
\begin{equation}
	\zeta^i_{0} (\vu{s}) = \mathcal{S}^i(\vu{s}) = -\sum_{l= 0}^\infty \sum_{m = -l}^l g^l Y_l^{m \ast}(\vu{s}) \int_{4\piup} \dd{\Omega^\prime} \frac{Y_l^m(\vu{s}^\prime) s_i^\prime}{\mus(\vu{s}^\prime)},
\end{equation}
which can be rewritten as
\begin{equation}\label{eq:zeta0_2}
	\zeta^i_{0} (\vu{s}) = \mathcal{S}^i(\vu{s}) = -\sum_{l= 0}^\infty \sum_{m = -l}^l g^l Y_l^{m \ast}(\vu{s}) H_{lm}^i,
\end{equation}
with
\begin{equation}\label{eq:Hlm}
	H_{lm}^i = \int_{4\piup} \dd{\Omega^\prime} \frac{Y_l^m(\vu{s}^\prime) s_i^\prime}{\mus(\vu{s}^\prime)},
\end{equation}
which is a function of $\mux$, $\muy$ and $\muz$.
We now replace Eq.~\eqref{eq:zeta0_2} in Eq.~\eqref{eq:zetan} with $n = 1$ to obtain
\begin{equation}
	\begin{split}
		\zeta^i_{1} (\vu{s}) &= \sum_{l= 0}^\infty \sum_{m = -l}^l g^{l} Y_l^{m \ast}(\vu{s}) \int_{4\piup} \dd{\Omega^\prime} \zeta_0^i (\vu{s}^\prime) Y_{l}^{m} (\vu{s}^\prime) \\
		&= -\sum_{l= 0}^\infty \sum_{m = -l}^l \sum_{l'= 0}^\infty \sum_{m' = -l'}^{l'} g^{l + l'} Y_l^{m \ast}(\vu{s}) H_{lm}^i \int_{4\piup} \dd{\Omega^\prime} Y_{l'}^{m' \ast}(\vu{s}^\prime)Y_{l}^{m} (\vu{s}^\prime)\\
		&= -\sum_{l= 0}^\infty \sum_{m = -l}^l g^{2l} Y_l^{m \ast}(\vu{s}) H_{lm}^i
	\end{split}
\end{equation}
where we have used the orthonormality relation for the spherical harmonics
\begin{equation}
	\int_{4\piup} \dd{\Omega} Y_l^{m \ast}(\vu{s})Y_{l'}^{m'} (\vu{s}) = \delta_{l, l'} \delta_{m, m'}.
\end{equation}

Similarly, it can be demonstrated by induction that
\begin{equation}\label{eq:zetan_2}
	\zeta^i_{n} (\vu{s}) = - \sum_{l= 0}^\infty \sum_{m = -l}^l g^{(n+1)l} Y_l^{m \ast}(\vu{s}) H_{lm}^i.
\end{equation}
Substituting this result into Eq.~\eqref{eq:zeta} returns
\begin{equation}
	\zeta^i(\vu{s}) = \sum_{n = 0}^\infty \zeta_n^i(\vu{s}) = -\sum_{l= 0}^\infty \sum_{m = -l}^l Y_l^{m \ast}(\vu{s}) H_{lm}^i g^l\sum_{n = 0}^\infty g^{nl},
\end{equation}
which can be expressed as
\begin{equation} \label{eq:zeta_2}
	\zeta^i(\vu{s}) = -\sum_{l= 0}^\infty \sum_{m = -l}^l \frac{g^l}{1 - g^l} Y_l^{m \ast}(\vu{s}) H_{lm}^i.
\end{equation}

Inserting Eq.~\eqref{eq:zeta_2} in Eq.~\eqref{eq:Dijgneq0} yields
\begin{equation} \label{eq:Dij_2}
	\begin{split}
		D_{ij} &= \delta_{ij} \eval{D_{ii}}_{g = 0} - \frac{v}{4\piup\lavg} \int_{4\piup} \frac{ \zeta^i(\vu{s})s_j}{\mus(\vu{s})} \dd{\Omega}\\
		&= \delta_{ij} \eval{D_{ii}}_{g = 0} + \frac{v}{4\piup\lavg} \sum_{l= 0}^\infty \sum_{m = -l}^l \frac{g^l}{1 - g^l} H_{lm}^i \int_{4\piup} \dd{\Omega} \frac{s_jY_l^{m \ast}(\vu{s})}{\mus(\vu{s})}\\
		&= \delta_{ij} \eval{D_{ii}}_{g = 0} + \frac{v}{4\piup\lavg} \sum_{l= 0}^\infty \sum_{m = -l}^l \frac{g^l}{1 - g^l} H_{lm}^i H_{lm}^{j, \ast}.
	\end{split}
\end{equation}

We can now simplify the expressions by exploiting the symmetries of spherical harmonics and of $\mus(\vu{s})$.
If we consider inversion $\vu{s} \to -\vu{s}$, we have
\begin{equation}
	Y_l^m(-\vu{s}) = (-1)^lY_l^m(\vu{s}), \quad \mus(-\vu{s}) = \mus(\vu{s}),
\end{equation}
and therefore
\begin{equation}
	H_{lm}^i = \int_{4\piup} \dd{\Omega} \frac{Y_l^m(\vu{s}) s_i}{\mus(\vu{s})} = \int_{4\piup} \dd{\Omega} \frac{Y_l^m(-\vu{s}) \left(-s_i\right)}{\mus(-\vu{s})} = - (-1)^l H_{lm}^i,
\end{equation}
which implies
\begin{equation}\label{eq:simm4}
	H_{2l, m}^i = 0 \quad \forall l.
\end{equation}

When we consider instead the transformation $(x, y, z)\to (-x, -y, z)$, corresponding to $\phi \to \phi + \piup$, we have
\begin{equation}
	Y_l^m(\theta,\phi + \piup) = (-1)^m Y_l^m(\theta,\phi), \quad \mus(\theta,\phi + \piup) = \mus(\theta,\phi),
\end{equation}
and for the $x$ and $y$ components of $H_{lm}$ we obtain
\begin{equation}
	\begin{split}
		H_{lm}^{x,y} &= \int_{4\piup} \dd{\Omega} \frac{Y_l^m(\theta,\phi) s_{x,y}(\theta,\phi)}{\mus(\theta,\phi )} \\
		&= \int_{4\piup} \dd{\Omega} \frac{Y_l^m(\theta,\phi + \piup) s_{x,y}(\theta,\phi + \piup)}{\mus(\theta,\phi + \piup)} = - (-1)^m H_{lm}^{x,y},
	\end{split}
\end{equation}
while for the $z$ component we have
\begin{equation}
	H_{lm}^{z} = (-1)^m H_{lm}^z,
\end{equation}
from which it follows that
\begin{equation} \label{eq:simm1}
	H_{2l, 2m+1}^z = H_{2l, 2m}^{x, y} = 0 \quad \forall l, m.
\end{equation}

If we instead consider reflection about the $y$ axis, corresponding to $\phi \to -\phi$, we have
\begin{equation}
	Y_l^m (\theta, -\phi) = Y_l^{m\ast} (\theta, \phi), \quad \mus(\theta, -\phi) = \mus(\theta, \phi),
\end{equation}
which gives
\begin{equation} \label{eq:simm2}
	H_{lm}^{x,z} = H_{lm}^{x,z\ast}, \quad H_{lm}^y = - H_{lm}^{y\ast} \quad \forall l, m,
\end{equation}
that is, the $x$ and $z$ components of $H_{lm}$ are real, while the $y$ component is imaginary.

Furthermore, since $Y_l^{-m}(\theta, \phi) = (-1)^m Y_l^{m \ast}(\theta, \phi)$, from Eq.~\eqref{eq:Hlm} we also find
\begin{equation} \label{eq:simm3}
	H_{l, -m}^i = (-1)^m H_{l, m}^{i\ast} \quad \forall l, m.
\end{equation}

We can now show that the off-diagonal elements of the diffusion tensor vanish for $g\neq0$.
By using Eq.~\eqref{eq:simm1}, it follows that
\begin{equation}
	H_{2l, m}^{x,y} H_{2l, m}^{z\ast} = 0 \quad \forall m,
\end{equation}
because for each $m$ one of the two factors is identically zero.
As for the tensor components $D_{xy}$ and $D_{yx}$, using Eqs.~\eqref{eq:simm4}, \eqref{eq:simm1}, \eqref{eq:simm2}, \eqref{eq:simm3} we can rewrite the sum over $m$ in Eq.~\eqref{eq:Dij_2} as
\begin{equation}
	\begin{split}
		\sum_{m = -2l - 1}^{2l +1} H_{2l+1, m}^x H_{2l+1, m}^{y\ast} &= \sum_{m = 0}^{l}( H_{2l+1, 2m+ 1}^x H_{2l+1, 2m+1}^{y\ast} + H_{2l+1, -2m-1}^x H_{2l+1, -2m-1}^{y\ast})\\
		&= \sum_{m = 0}^{l}( H_{2l+1, 2m+ 1}^x H_{2l+1, 2m+1}^{y\ast} + H_{2l+1, 2m+1}^{x\ast} H_{2l+1, 2m+1}^{y})\\
		&=\sum_{m = 0}^{l}2\Re( H_{2l+1, 2m+ 1}^x H_{2l+1, 2m+1}^{y\ast} ) = 0.
	\end{split}
\end{equation}

Finally, this allows us to write the diffusion tensor components as
\begin{equation} \label{eq:Dij_g}
	D_{ij} = \delta_{ij} \eval{D_{ii}}_{g = 0} + \frac{v \delta_{ij}}{4\piup\lavg} \sum_{n= 0}^\infty \sum_{m = -2n -1}^{2n + 1} \frac{g^{2n + 1}}{1 - g^{2n+1}} \abs{H_{2n+1,m}^i}^2.
\end{equation}

Using the symmetry $\abs{H^i_{l,-m}}^2 = \abs{H^i_{l,m}}^2$ together with the selection rules derived above, only a subset of the $m$ indices contributes to the sum.
In particular, for odd angular orders $l=2n+1$, the non-vanishing terms can be grouped pairwise, allowing the expression for the diagonal diffusion tensor elements to be written in the more compact form used in the main text
\begin{equation} \label{eq:Di_app}
	D_i = \eval{D_i}_{g = 0} + \frac{v}{4\piup \lavg}\sum_{n = 0}^\infty \frac{g^{2n + 1}}{1 - g^{2n+1}} \sum_{m=-n}^n \abs{H^i_{2n+1, 2m}}^2.
\end{equation}

In the isotropic limit, Eq.~\eqref{eq:Di_app} reduces to the expected form $\lim_{\mus(\vu{s}) \to \mus} D_i = v/(3\mus) + vg/[3\mus(1-g)] = v/[3\mus(1-g)]$.

\section{Direction-dependent persistence length} \label{appB}
In the presence of a scalar single-scattering asymmetry factor $g$, directional correlations persist over multiple scattering events and modify the effective length scale entering boundary conditions and source-depth placement.
We introduce a direction-dependent \emph{persistence length} $\lambda(\vu{s})$, which provides the exact length scale for a Henyey--Greenstein phase function with scalar $g$.

Let $\vu{s}_0,\vu{s}_1,\ldots$ denote successive propagation directions of a random walker undergoing scattering with phase function $p(\vu{s} \cdot \vu{s}^\prime)$.
We define the conditional persistence vector
\begin{equation}
	\boldsymbol{\Lambda}(\vu{s})
	=
	\left\langle
	\sum_{n=0}^{\infty}
	\ls(\vu{s}_n) \vu{s}_n
	\middle|
	\vu{s}_0=\vu{s}
	\right\rangle,
\end{equation}
where $\ls(\vu{s})=1/\mus(\vu{s})$ is the direction-dependent mean free path.

This quantity satisfies the renewal equation
\begin{equation}
	\boldsymbol{\Lambda}(\vu{s})
	=
	\ls(\vu{s}) \vu{s}
	+
	\int_{4\piup}
	p(\vu{s} \cdot \vu{s}^\prime)
	\boldsymbol{\Lambda}(\vu{s}^\prime)
	\dd{\Omega^\prime}.
	\label{eq:Lambda_resolvent}
\end{equation}
Introducing the scattering operator
\begin{equation}
	(Kf)(\vu{s}) = \int_{4\piup} p(\vu{s}\cdot\vu{s}') f(\vu{s}') \dd{\Omega'},
\end{equation}
Eq.~\eqref{eq:Lambda_resolvent} can be written as $(\mathbb{I}-K)\boldsymbol{\Lambda}=\ls(\vu{s})\vu{s}$.

The scalar persistence length relevant for boundary conditions is obtained by projecting $\boldsymbol{\Lambda}$ along the initial propagation direction,
\begin{equation}
	\lambda(\vu{s})
	=
	\vu{s}\cdot\boldsymbol{\Lambda}(\vu{s})
	=
	\left\langle
	\sum_{n=0}^{\infty}
	\ls(\vu{s}_n)
	(\vu{s}_n \cdot \vu{s})
	\middle|
	\vu{s}_0=\vu{s}
	\right\rangle.
	\label{eq:lambda_def}
\end{equation}

For a scalar Henyey--Greenstein phase function, the scattering operator is diagonal in spherical harmonics (Eq.~\eqref{eq:pHG}).
As a consequence, the resolvent $(\mathbb{I}-K)^{-1}$ introduces factors $(1-g^{l})^{-1}$ in each angular momentum channel.

Since $\ls(\vu{s}) \vu{s}$ is odd under inversion $\vu{s}\to-\vu{s}$, only odd Legendre orders contribute to the solution.
Using the spherical-harmonic addition theorem, the persistence length can be written directly in terms of Legendre polynomials as
\begin{equation}
	\lambda(\vu{s})
	=
	\sum_{n=0}^{\infty}\frac{4n+3}{4\piup}\frac{1}{1-g^{2n+1}}
	\int_{4\piup}\frac{\vu{s} \cdot \vu{s}^\prime P_{2n+1}(\vu{s} \cdot \vu{s}^\prime)}{\mus(\vu{s}^\prime)}\dd{\Omega^\prime}.
	\label{eq:persistence_length}
\end{equation}

Notably, Eq.~\eqref{eq:persistence_length} reduces to the expected limits $\lim_{g \to 0} \lambda(\vu{s})=\ls(\vu{s})$ and $\lim_{\mus(\vu{s}) \to \mus} \lambda(\vu{s})=\ls/(1-g)=\lt$.

\section{Derivation of the boundary-condition integral \texorpdfstring{\boldmath{$\Yze$}}{Y} with Fresnel reflections} \label{appC}
In this appendix we derive the boundary-condition integral $\Yze$ entering the extrapolated length $\ze$ in the presence of Fresnel reflections at the slab boundaries.
The derivation closely follows Appendix~\ref{appA}; therefore, only the steps that differ from the bulk calculation are reported.

The boundary-condition integral $\Yze$ is defined as
\begin{equation} \label{eq:Y_appC_def}
	\Yze = \frac{1}{4\piup\lavg} \int_{\Omega_{\text{up}}}\left(\frac{s_z^2}{\mus^2(\vu{s})}-\frac{s_z \zeta^z(\vu{s})}{\mus(\vu{s})}\right)R(\theta)\dd{\Omega},
\end{equation}
where $\Omega_{\text{up}}$ denotes the upper hemisphere ($s_z>0$) and $R(\theta)$ is the Fresnel reflection coefficient defined in Eq.~\eqref{eq:Rfresnel}.
The auxiliary function $\zeta^z(\vu{s})$ is the $z$ component of Eq.~\eqref{eq:zeta_2}, since its defining integral equation involves only bulk scattering properties and is unaffected by boundary reflections.

Substituting the expression for $\zeta^z(\vu{s})$ into Eq.~\eqref{eq:Y_appC_def} yields
\begin{equation} \label{eq:Y_appC_insert}
	\Yze = \frac{1}{4\piup\lavg} \int_{\Omega_{\text{up}}} \frac{s_z^2}{\mus^2(\vu{s})} R(\theta)\dd{\Omega} + \frac{1}{4\piup\lavg} \sum_{l=0}^{\infty} \sum_{m=-l}^{l} \frac{g^l}{1-g^l}H_{lm}^z \int_{\Omega_{\text{up}}} \frac{s_z}{\mus(\vu{s})} Y_l^{m\ast}(\vu{s}) R(\theta)\dd{\Omega}.
\end{equation}

Because the angular integration is restricted to the upper hemisphere, the orthonormality relations of the spherical harmonics over $4\piup$ can no longer be exploited.
It is therefore convenient to introduce the hemisphere-weighted projection coefficient
\begin{equation} \label{eq:Htilde_appC}
	\widetilde{H}_{lm}^z = \int_{\Omega_{\text{up}}}\frac{s_z}{\mus(\vu{s})} Y_l^{m\ast}(\vu{s}) R(\theta)\dd{\Omega}.
\end{equation}

Using this definition, Eq.~\eqref{eq:Y_appC_insert} can be rewritten as
\begin{equation} \label{eq:Y_appC_general}
	\Yze = \eval{\Yze}_{g = 0} + \frac{1}{4\piup\lavg}\sum_{l=0}^{\infty} \sum_{m=-l}^{l} \frac{g^l}{1-g^l}H_{lm}^z\widetilde H_{lm}^z,
\end{equation}
where the isotropic-scattering contribution $\eval{\Yze}_{g=0}$ is given by Eq.~\eqref{eq:Y}.

In full analogy with the selection rules derived in Appendix~\ref{appA}, the sum is restricted to odd angular orders $l=2n+1$.
The final expression for $\Yze$ therefore becomes
\begin{equation} \label{eq:Y_appC_final}
	\Yze = \eval{\Yze}_{g = 0} + \frac{1}{4\piup\lavg} \sum_{n=0}^{\infty} \frac{g^{2n+1}}{1-g^{2n+1}} \sum_{m=-n}^n H_{2n+1,m}^z \widetilde H_{2n+1,m}^z.
\end{equation}

In the isotropic limit, $\ze$ tends to the standard form reported in Ref.~\cite{martelli2022light}, as expected.
	
\end{document}